\def\year{2020}\relax
\documentclass[letterpaper]{article} 
\usepackage{aaai20}  
\usepackage{times}  
\usepackage{helvet} 
\usepackage{courier}  
\usepackage[hyphens]{url}  
\usepackage{graphicx} 
\usepackage{graphicx}  
\usepackage{graphicx,color}
\usepackage{amsmath, array}
\usepackage{placeins}
\usepackage{changepage}
\usepackage{longtable}
\usepackage{multirow}
\usepackage{siunitx}
\usepackage{supertabular}
\usepackage{enumitem}
 \usepackage{textcomp} 
 \usepackage{subfigure}
 \usepackage{xcolor}

\title{A Dataset of State-Censored Tweets}

\author{Tuğrulcan Elmas\\
EPFL\\
tugrulcan.elmas@epfl.ch}
\author{Rebekah Overdorf\\
EPFL\\
rebekah.overdorf@epfl.ch}
\author{Tuğrulcan Elmas, Rebekah Overdorf, Karl Aberer\\
EPFL\\
karl.aberer@epfl.ch}
\author{Tuğrulcan Elmas, Rebekah Overdorf, Karl Aberer\\
EPFL\\
tugrulcan.elmas@epfl.ch, rebekah.overdorf@epfl.ch, karl.aberer@epfl.ch}

\begin{document}

\newcommand{\Secref}[1]{Section~\ref{#1}}
\newcommand{\Figref}[1]{Figure~\ref{#1}}

\newcommand{\descr}[1]{{\bigskip\noindent\textbf{#1}}}
\newcommand{\descrplain}[1]{{\smallskip\noindent\textbf{#1}}}

\newcommand{\attr}[1]{#1}

\newcommand{\bekah}[1]{{\color{blue}BO: #1}}


\maketitle
\begin{abstract}

Many governments impose traditional censorship methods on social media platforms. Instead of removing it completely, many social media companies, including Twitter, only withhold the content from the requesting country. This makes such content still accessible outside of the censored region, allowing for an excellent setting in which to study government censorship on social media. We mine such content using the Internet Archive's Twitter Stream Grab. We release a dataset of 583,437 tweets by 155,715 users that were censored between 2012-2020 July. We also release 4,301 accounts that were censored in their entirety. Additionally, we release a set of 22,083,759 supplemental tweets made up of all tweets by users with at least one censored tweet as well as instances of other users retweeting the censored user. We provide an exploratory analysis of this dataset. Our dataset will not only aid in the study of government censorship but will also aid in studying hate speech detection and the effect of censorship on social media users. The dataset is publicly available at https://doi.org/10.5281/zenodo.4439509
\end{abstract}
\section{Introduction}

\hyphenation{BuzzFeed}

While there are many disagreements and debates about the extent to which a government should be able to censor and which content should be allowed to be censored, many governments impose some limitations on content that can be distributed within their borders. The specific reasons for censorship vary widely depending on the type of censored content and the context of the censor, but the overall goal is always the same: to suppress speech and information. 

Traditional Internet censorship works by filtering by IP address, DNS filtering, or similar means. The rise of social media platforms has lessened the impact of such techniques. As such, modern censors often instead issue take-down requests to social media platforms. As a practice, Twitter does not normally remove content that does not violate the terms of service but instead withholds the content from that country if the respective government sends a legal request~\cite{twitter_withheld}. As such, these tweets are still accessible on the platform from outside of the censored region, which provides a setting to study the censorship policies of certain governments. In this paper, we release an extensive dataset of tweets that were withheld between July 2012 and July 2020. Aside from an initiative by BuzzFeed journalists who collected a smaller dataset of users that were observed to be censored between October 2017 and January 2018, such dataset has never been collected and released. We also release uncensored tweets of users which are censored at least once. As such, our dataset will help not only in studying government censorship itself but also the effect of censorship on users and the public reaction to the censored tweets. 


The main characteristics of the datasets are as follows:

\begin{enumerate}
    \item 583,437 censored tweets from 155,715 unique users.
    \item 4,301 censored accounts.
    \item 22,083,759 additional tweets from users who posted at least one censored tweet and retweets of those tweets.
\end{enumerate}

We made the dataset available at Zenodo: https://doi.org/10.5281/zenodo.4439509. The dataset only consists of tweet ids and user ids in order to comply with Twitter Terms of Service. For the documentation and the code to reproduce the pipeline please refer to https://github.com/tugrulz/CensoredTweets.

This paper is structured as follows. \Secref{sec:related} discusses the related work. \Secref{sec:collection} describes the collection pipeline. \Secref{sec:exploratory} provides an exploratory analysis of the dataset. \Secref{sec:usecases} describes the possible use cases of the dataset. Finally, \Secref{sec:ethics} discusses the caveats: biases of the dataset, ethical considerations, and compliance with FAIR principles.

\section{Background and Related Work}
\label{sec:related}

Traditional Internet censorship blocks access to an entire website within a country by filtering by IP address, DNS filtering, or similar means. Such methods, however, are heavy-handed in the current web ecosystem in which a lot of content is hosted under the same domain. That is, if a government wants to block one Wikipedia page, it must block every Wikipedia page from being accessible within its borders. Such unintended blockings, so-called ``casualties,'' are tolerated by some censors, who balance the tradeoff between the amount of benign content and targeted content being blocked. Some countries have gone so far as to ban entire popular websites based on some content. Turkey famously blocked Wikipedia from 2017 through January 2020~\cite{bbc1} in response to Wikipedia articles that it deemed critical of the government. Several countries, including Turkey, China, and Kazakhstan have intermittently blocked WordPress, which about 39\% of the Internet is built on\footnote{\url{https://kinsta.com/wordpress-market-share/}}. 

To study this type of censorship, researchers, activists, and organizations complete measurement studies to determine which websites are blocked or available in different regions. Primarily, this tracking relies on volunteers running software from inside the censor, such as the tool provided by the Open Observatory of Network Interference (OONI)\footnote{\url{https://ooni.org/}}. 

The rise of social media platforms further complicates censorship techniques. In order to censor a single post, or even a group or profile, the number of casualties is by necessity very high, up to every social media profile and page. In response, some censors now request that social media companies remove content, or at least block certain content within their borders. To a large extent, social media companies comply with these requests. For example, in 2019, Reddit, which annually publishes a report~\cite{reddit_transparency_report} on takedown requests, received 110 requests to restrict or remove content and complied with 41. In the first half of 2020, Twitter received 42.2k requests for takedowns and complied with 31.2\%. Twitter states that they have censored 3,215 accounts and 28,370 individual tweets between 2012 and July 2020. They also removed 97,987 accounts altogether from the platform after a legal request instead of censoring~\cite{twitter_transparency_report}. Measuring this type of targeted censorship requires new data and new methodologies. This paper takes the first step in this direction by providing a dataset of regionally censored tweets. 


To the best of our knowledge, only one similar dataset has been publicly shared: journalists at BuzzFeed manually curated and shared a list of 1,714 censored accounts~\cite{buzzfeed}. This dataset covers users whose entire profile was observed to be censored between October 2017 and January 2018 but does not cover individual censored tweets. The Lumen Database\footnote{\url{https://www.lumendatabase.org}} stores the legal demands to censor content sent to Twitter in pdf format which includes the court order and the URLs of the censored tweets. Although the database is open to the public, one has to send a request for every legal demand, solve a captcha, download, and open a pdf to access a single censored tweet, which makes the database inaccessible and not interoperable. Although not publicly available, using a dataset of censored tweets from October 2014 to January 2015 in Turkey, previous research found that most were political and critical of the government~\cite{tanash2015known,tanash2016detecting} and that Twitter underreports the censored content. Subsequent work reported a decline in government censorship~\cite{tanash2017decline}. Additionally, ~\cite{varol2016spatiotemporal} analyzed 100,000 censored tweets between 2013 and 2015 and reported that the censorship does not prevent the content from reaching a broader audience. There exists no other robust, data-driven study of censored Twitter content to the best of our knowledge. This is likely due in part to the difficulty of collecting a dataset of censored tweets and accounts. Our dataset aims to facilitate such work. 
\section{Data Collection}
\label{sec:collection}

\begin{figure*}
    \centering
    \includegraphics[width = \textwidth]{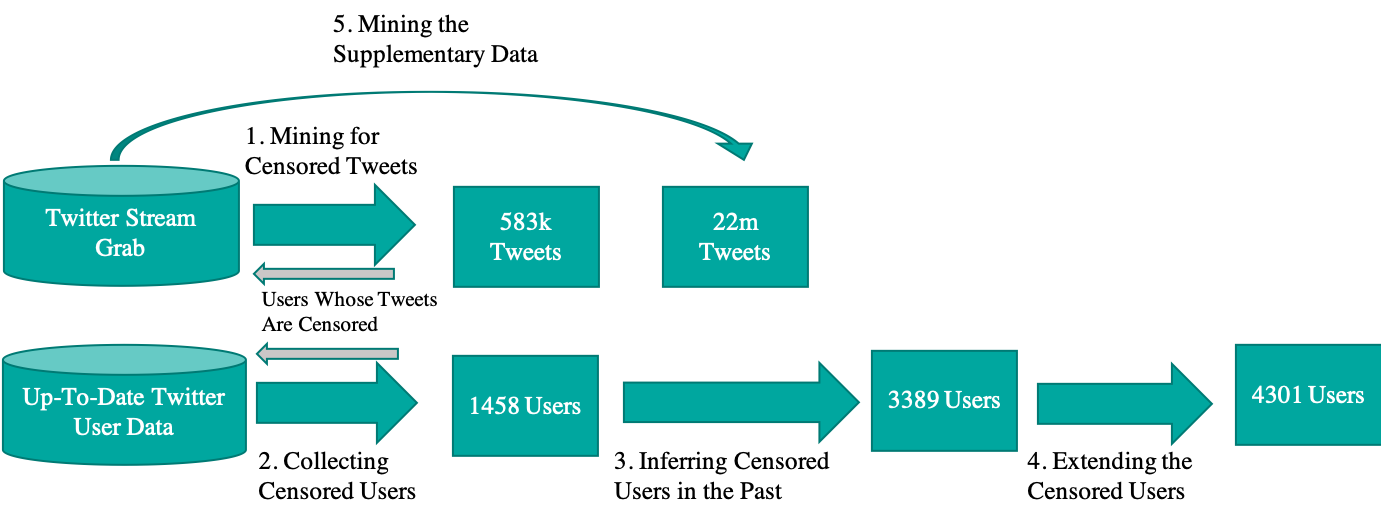}
    \caption{Overview of the data collection process. We first mined the entire Twitter Stream Grab (published by the Internet Archive), which consists of a 1\% sample of all tweets, and selected the tweets that were denoted as being censored. This resulted in 583k censored tweets. We next created a list of accounts that had at least one censored tweet and determined, via the user lookup endpoint, which accounts were fully censored and which only had selective tweets censored. We then inferred which of these accounts were censored in the past (see Section 3.3). We next extended the censored users dataset by exploiting the users' social connections. Finally, we mined the Twitter Stream Grab in order to collect the other (non-censored) tweets of users who had been censored at least once.}
    \label{fig:summary}
\end{figure*}

This section describes the process by which we collect censored tweets and users, which is summarized in Figure \ref{fig:summary}. In brief, data retrieved via the Twitter API is structured as follows: tweets are instantiated in a Tweet object, which includes, among other attributes, another object that instantiates the tweet's author (a User object). If the tweet is a retweet, it also includes the tweet object of the retweeted tweet, which in turn includes the user object of the retweeted user. If a tweet is censored, its tweet object includes a ``withheld\_in\_countries'' field which features the list of countries the tweet is censored in. If an entire profile is censored, the User object includes the same field. In order to build a dataset of censored content, our objective is to find all tweet and user objects with a ``withheld\_in\_countries'' field. 

For this objective, we first mine the Twitter Stream Grab, which is a dataset of 1\% sample of all tweets between 2011 and 2020 July (\Secref{sec:mining}). This dataset does not provide information on whether the entire profile is censored. As such, we collect the censored users from the up-to-date Twitter data using the Twitter User Lookup API endpoint (\Secref{sec:uptodate}). We infer if a user was censored in the past by a simple procedure we come up with (\Secref{sec:infer}). We extend the dataset of censored users exploiting their social connections (\Secref{sec:extend}). We lastly mine a supplementary dataset consisting of non-censored tweets by users who were censored at least once (\Secref{sec:supplementary}). The whole collection process is described in Figure \ref{fig:summary}. We now describe the process in detail.

\subsection{Mining The Twitter Stream Grab For Censored Tweets}
\label{sec:mining}

Twitter features an API endpoint that provides 1\% of all tweets posted in real-time, sampled randomly. The Internet Archive collects and publicly publishes all data in this sample starting from September 2011. They name this dataset the \emph{Twitter Stream Grab}~\cite{team2020archive}. At the time of this analysis, the dataset consisted of tweets through June 2020. Although the dataset is publicly available, only a few studies~\cite{tekumalla2020mining,elmas2020misleading,elmas2020power} have mined the entire dataset due to the cumbersome process of storing and efficiently processing the data. We mine all the tweets in this dataset to find tweets and users with the ``withheld\_in\_countries'' field and, thus, find the censored tweets and users.

We process all the tweets between September 2011 and June 2020 in this dataset and retain those with the ``withheld\_in\_countries'' field. We found 583,437 tweets from 155,715 users in total. Of these, 378,286 were retweets. The tweet ids in this dataset can be found in \texttt{{tweets.csv}} and their authors' user ids can be found in \texttt{all\_users.csv}.

\subsection{Collecting Censored Users From Up-To-Date Twitter User Data}

\label{sec:uptodate}

The user objects in the Twitter Stream Grab data do not include a ``withheld\_in\_countries'' field due to the limits of the API endpoint it uses. Twitter provides this field in the data returned by the User Lookup endpoint. As the Internet Archive does not store the past data provided by this endpoint, we collect censored users by communicating with this end-point in December 2020. Precisely, from the list of users with at least one censored tweet, we selected the users whose entire profile was censored as of December 2020. One caveat is that Twitter only provides the data of accounts that still exist on the platform. Out of the 155,715 candidate users, 114,800 (73.7\%) still existed on the platform as of December 2020. Of the remaining 40,915 users, 62.2\% were suspended by Twitter. Of the users that still existed, we found that 1,458 had their entire profile censored.

\subsection{Inferring Censored Users In The Past}

\label{sec:infer}

For the users that have at least one censored tweet and who did not exist on the platform as of December 2020, we do not know if their entire profile was censored. Additionally, for those users that have at least one censored tweet, but whose accounts were not found to be censored after retrieving the up-to-date content, we can not know if their entire profile was censored in the past. We developed a strategy to infer which users had their entire profile censored at some point in the past. We observe that some users’ censored tweets were in fact retweets, but the tweets that were retweeted by those users were not censored. This is either because the legal request was sent for only the retweet and not the original tweet or all the tweets (including the retweets) of the same account were censored automatically because the account itself was censored. We believe the former is unlikely because censoring the retweet does not block the visibility of the content, which is the goal. Additionally, we observed retweets that were unlikely candidates for censorship such as those posted by \textit{@jack}, Twitter’s founder. Thus, when we observed a censored retweet of a non-censored tweet, we assume the retweeting account was censored as a whole. Using this reasoning, we found 3,063 censored accounts of which 1,531 were still on the platform and 319 were no longer censored. This method missed 326 users who were actually censored, achieving 77.6\% recall. This increased the number of users to 3,389. We provide the list of users found to be censored only by this procedure in \texttt{usersinferred.csv}.

\subsection{Extending The Censored Users}

\label{sec:extend}

Up to this point, the number of users whose entire profile was censored is 3,389, which is bigger than the dataset created by BuzzFeed which consisted of 1,714. However, when we evaluate our datasets' recall on BuzzFeed's dataset, we found that we only captured 65\% of the users in that dataset. To increase recall, we extend our dataset by exploiting the connections of the list of censored users we have. Precisely, we collect the friends of the accounts in our dataset whose tweets were censored and the Twitter lists they are added to. For the latter case, we then collect the members of these lists. We collect 3,233,554 friends and the members of 70,969 lists in total. We found 494 censored users, increasing the recall to 80.3\% when evaluated using the BuzzFeed dataset. We merge the extended dataset with the BuzzFeed dataset which is separately stored in \texttt{buzzfeed\_users.csv}. We collected 4,301 users whose profile was entirely censored in total. The resulting dataset can be found in \texttt{users.csv}.

\subsection{Mining The Supplementary Data}

\label{sec:supplementary}

Finally, we mined the Twitter Stream Grab for the tweets of users with at least one censored tweet and retweets of those users by others. We found 22,083,759 such tweets. These tweets will include non-censored tweets and might serve as negative samples for studies that would consider the censored tweets as positive samples. They can be found in \texttt{supplement.csv}.
\section{Exploratory Analysis}
\label{sec:exploratory}

\begin{figure*}[ht]
    \centering
    \includegraphics[width = 2\columnwidth]{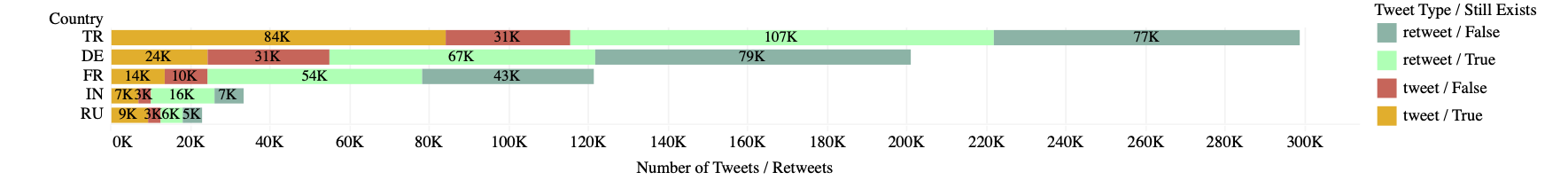}
    \caption{The statistics of the tweets and retweets with respect to countries and the existence of tweets on the platform by December 2020. }
    \label{fig:fivecountries}
\end{figure*}

We first begin by reporting the statistics of the tweets per their current status. Of the 583,437 censored tweets, 328,873 (56.3\%) were still present (not deleted or removed from) the platform as of December 2020. Of those that remained, 4,716 tweets were no longer censored, nor were their authors. Of those that were still censored, 154,572 of the tweets were posted by accounts that were censored as a whole, and 168,234 tweets were posted by accounts that were not censored. 

We continue with the per-country analysis of the censorship actions. Although the dataset features 13 different countries, we found that 572,095 tweets (98\%) were censored in only five countries: Turkey, Germany, France, India, and Russia. Figure \ref{fig:fivecountries} shows the statistics with respect to these countries.

\begin{figure*}[t]
\centering     
\subfigure{\label{fig:a}\includegraphics[width=0.45\textwidth]{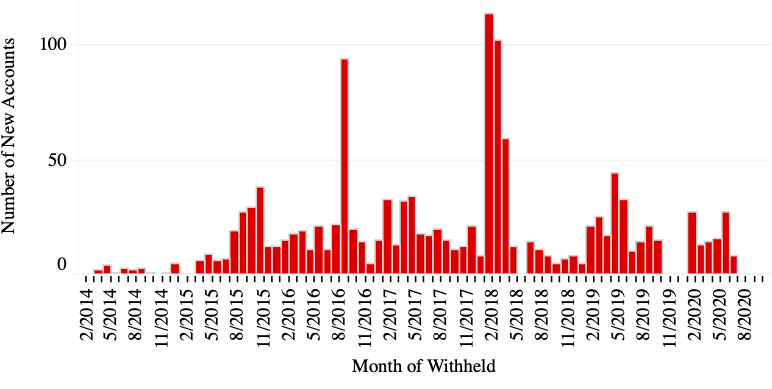}} \hfill	
\subfigure{\label{fig:b}\includegraphics[width=0.45\textwidth]{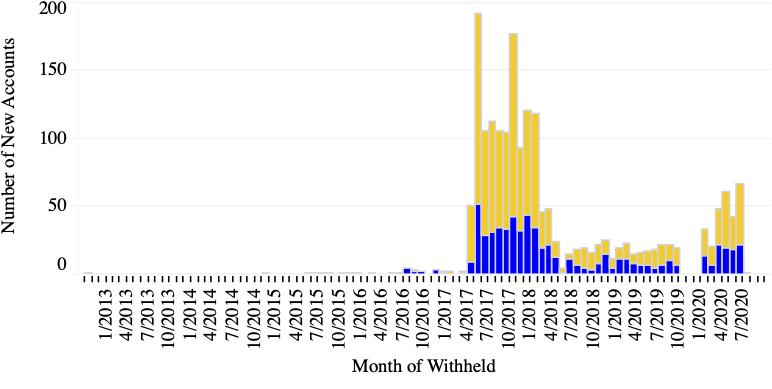}}
\\\vspace{-15pt}
\subfigure{\label{fig:c}\includegraphics[width=0.45\textwidth]{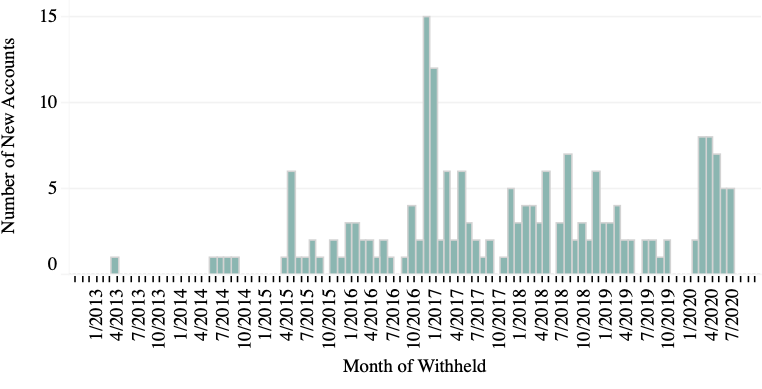}} \hfill
\subfigure{\label{fig:d}\includegraphics[width=0.45\textwidth]{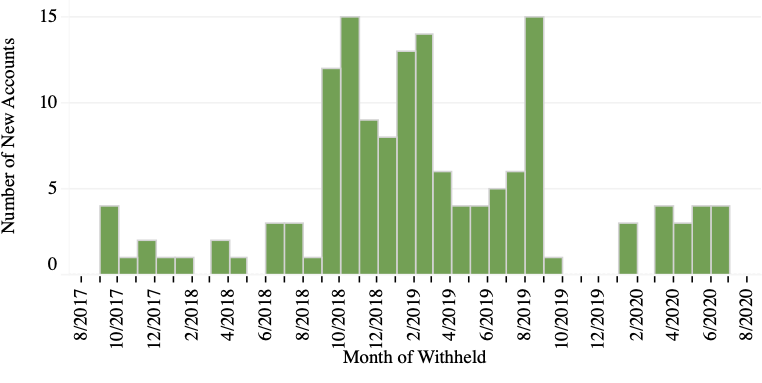}}

\caption{The temporal activity of the censored tweets, measured by the number of users censored for the first time, with respect to countries. The censorship actions by Turkey, Russia, and India appear to follow the domestic politics and regional conflicts involving those countries. The censorship actions by Germany and France appear to follow the account creation date of the censored users. We did not identify any major event around the time of the censorship actions. }

\label{fig:temporal}

\end{figure*}

To better understand what this dataset consists of, we perform a basic temporal and a topical analysis for each of these five countries. Researchers can use this analysis as a starting point. We perform the topical analysis by computing the most popular hashtags, mentions, URLs and then observing those entities. We measure the popularity by the number of unique accounts mentioning each entity. We use this metric instead of the tweet count in order to account for the same users using the same entities over and over. We do not include the retweets in this analysis. Table ~\ref{tab:entities} shows those entities and their popularity. We additionally report the most frequent tweeting languages (measured by the number of tweets) and most reported locations (measured by the number of users) with respect to the censoring countries in Table ~\ref{tab:locations}. We perform the temporal analysis by computing the number of users censored tweet-wise or account-wise for the first time per month in Figure~\ref{fig:temporal}. We now briefly describe our findings.

\definecolor{green_dot}{HTML}{15534c}
\definecolor{yellow_x}{HTML}{a57d50}

\begin{table*}[]
\centering
\label{tab:entities}
\caption{Entities used per number of distinct users with respect to the countries they are censored in.}
\resizebox{\textwidth}{!}{%
\begin{tabular}{l|l|l|l}
Country & Hashtags                                                                                                                       & Mentions                                                                                                                             & URLs                                                                                                                                           \\ \hline
Turkey  & \begin{tabular}[c]{@{}l@{}}Afrin (99), YPG (83), Turkey (77), \\ Rojava (63), Efrin (62)\end{tabular}                          & \begin{tabular}[c]{@{}l@{}}YouTube (96), hrw (58), Enes\_Kanter (58),\\  amnesty (52), UN (43)\end{tabular}                          & \begin{tabular}[c]{@{}l@{}}pscp.tv (60),  www.tr724.com (48), www.shaber3.com (39),\\  anfturkce.com (33), www.hawarnews.com (28)\end{tabular} \\ \hline
Germany & \begin{tabular}[c]{@{}l@{}}Antifa (17), WhiteGenocide (16), MAGA (15), \\ ThursdayThoughts (14), Merkel (14)\end{tabular}      & \begin{tabular}[c]{@{}l@{}}realDonaldTrump (77), YouTube (53), POTUS (30), \\ CNN (26), FoxNews (25)\end{tabular}                    & \begin{tabular}[c]{@{}l@{}}gab.ai (19), www.welt.de (14), www.dailymail.co.uk (12), \\ www.rt.com (11), www.bbc.com (11)\end{tabular}          \\ \hline
France  & \begin{tabular}[c]{@{}l@{}}WhiteGenocide (11), ThursdayThoughts (7), \\ Islam (7), AltRight (7), WW2 (6)\end{tabular}          & \begin{tabular}[c]{@{}l@{}}YouTube (25), realDonaldTrump (24), Nature\_and\_Race (10), \\ CNN (10), RichardBSpencer (9)\end{tabular} & \begin{tabular}[c]{@{}l@{}}www.dailymail.co.uk (8), www.bbc.com (6), gab.ai (5), \\ www.rt.com (4), www.breitbart.com (4)\end{tabular}         \\ \hline
India   & \begin{tabular}[c]{@{}l@{}}Kashmir (53), Pakistan (20), India (15), \\ KashmirBleeds (9), FreeKashmir (9)\end{tabular}         & \begin{tabular}[c]{@{}l@{}}ImranKhanPTI (18), UN (13), OfficialDGISPR (11), \\ peaceforchange (9), Twitter (8)\end{tabular}          & \begin{tabular}[c]{@{}l@{}}flwrs.com (7), www.pscp.tv (4), tribune.com.pk (4), \\ www.theguardian.com (3), www.nytimes.com (3)\end{tabular}    \\ \hline
Russia  & \begin{tabular}[c]{@{}l@{}}Khilafah (12), Pakistan (11), FreeNaveedButt (10), \\ Syria (9), ReturnTheKhilafah (9)\end{tabular} & \begin{tabular}[c]{@{}l@{}}Youtube (10), poroshenko (3), vladydady1 (2), \\ viking\_inc\_ (2), sevastopukr (2)\end{tabular}          & \begin{tabular}[c]{@{}l@{}}www.hizb-ut-tahrir.info (15), pravvysektor.info (6), \\ www.guardian.com (5), vk.cc (5), vk.com (4)\end{tabular}   

\end{tabular}
}
\end{table*}

\begin{table}[]
\centering
\label{tab:locations}
\caption{The most frequent self reported locations (measured by number of users) and most frequent tweeting languages (measured by number of tweets) with respect to the countries they are censored in.}
\resizebox{\columnwidth}{!}{%
\begin{tabular}{l|l|l}
Country & Locations                                                                                                                     & Languages                                                                                          \\ \hline
Turkey  & \begin{tabular}[c]{@{}l@{}}United States (482), Germany (470), \\ İstanbul (297), Turkey (274), Kürdistan (204)\end{tabular}  & \begin{tabular}[c]{@{}l@{}}tr (73959), en (26101), de (1378), \\ fr (1237), ar (1115)\end{tabular} \\ \hline
Germany & \begin{tabular}[c]{@{}l@{}}United States (3544), Texas (692), \\ Florida (608), California (519), Portugal (341)\end{tabular} & \begin{tabular}[c]{@{}l@{}}en (33742), tr (3999), de (3160), \\ es (2327), ja (1050)\end{tabular}  \\ \hline
France  & \begin{tabular}[c]{@{}l@{}}United States (2817), Texas (517), \\ Florida (470), California (400), Portugal (337)\end{tabular} & \begin{tabular}[c]{@{}l@{}}en (18526), fr (632), ja (527), \\ es (483), fi (263)\end{tabular}      \\ \hline
India   & \begin{tabular}[c]{@{}l@{}}Pakistan (827), Lahore (389), Karachi (319),\\ Islamabad (292), Punjab (235)\end{tabular}          & \begin{tabular}[c]{@{}l@{}}en (1719), ur (2262), ar (252),\\ hi (73), in (48)\end{tabular}         \\ \hline
Russia  & \begin{tabular}[c]{@{}l@{}}Ukraine (217), Indonesia (47), Istanbul (30),\\ Ankara (28), Turkey (20)\end{tabular}              & \begin{tabular}[c]{@{}l@{}}en (3968), uk (2441), tr (1166),\\ ru (1030), ar (880)\end{tabular}    

\end{tabular}
}
\end{table}

We found that censored users are mostly based in a country other than they are censored in, beyond the reach of the law enforcement of the censoring country. In the case of Turkey, the users appear likely to be people who have emigrated to, e.g., Germany or The United States, as the users based in those countries primarily tweet in Turkish. For India and Russia, censored users are mostly based in neighboring and/or hostile countries (such as Ukraine and Pakistan). They might be the locals of those countries as they mostly tweet with the local language (Ukrainian and Urdu) of the countries they are based in. For the case of Germany and France, it appears that most of the censored users are also residents of foreign countries. However, these countries of residence (e.g., The United States and Portugal) are not neighboring and/or hostile.

\begin{figure}[ht]
    \centering
    \includegraphics[width=\columnwidth]{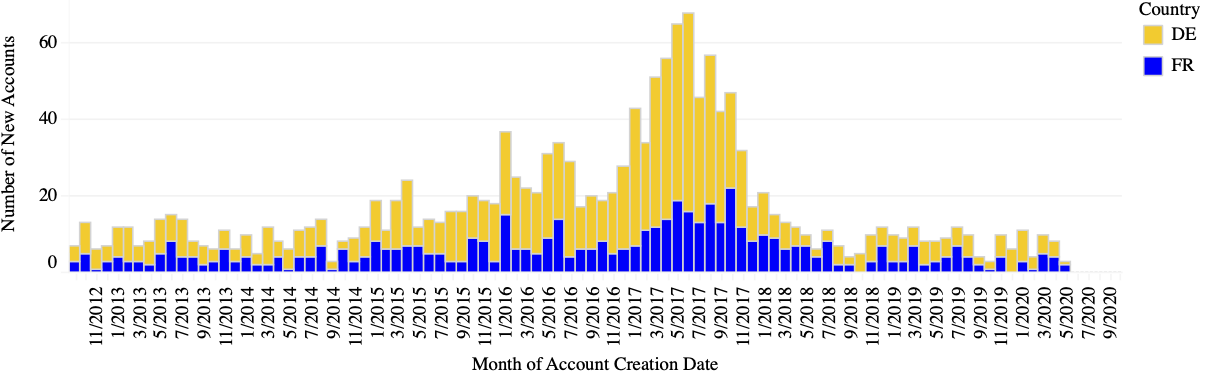}
    \caption{Account creation dates of users censored at least once by Germany and/or France. We observe an increasing trend of new accounts in late 2016 and 2017. The same trend can be observed during the period of censorship depicted in Figure~\ref{fig:temporal}}
    \label{fig:registration}
\end{figure}

The temporal activity of the censorship actions appeared to follow the domestic politics and regional conflicts involving Turkey, Russia, and India. This is further corroborated by the most popular hashtags. The biggest spike of users censored in Turkey was during Operation Olive Branch in which the Turkish army captured Afrin from YPG, one of the combatants of the Syrian Civil War based in Rojava, which is recognized as a terrorist organization by Turkey~\cite{dw}. The most popular hashtags among censored users were all related to this event. In the case of India, the most popular hashtags were about the Kashmir dispute. We could not identify any major event around the time of the spike in the number of censored accounts, but the dispute is still ongoing. The biggest spike of users censored in Russia coincided with the series of persecution of Hizbut Tahrir members, designated as a terrorist group by Russia~\cite{imrussia}. The accounts censored in Germany and France appear to be promoting extreme-right content since the frequent hashtags contained themes of white supremacy and Islamophobia. We could not identify any major event related to the spike of censorship observed in these countries. However, we found that the accounts censored were also newly registered to the platform when they were censored. 11.2\% of users censored by Germany and 9.1\% of users censored by France were first censored within one month of the account creation, while this is only 4.6\% for Turkey, 1\% for Russia, and 0\% for India. Figure \ref{fig:registration} shows that the creation of the accounts censored in Germany and France followed a similar trend to the time of censorship of the accounts in these two countries. Precisely, there was a surge of new accounts in late 2016 and 2017 which being to be censored in mid-2017. This might be due to the German and French governments' reaction to the many far-right accounts that started to actively campaign on Twitter in late 2016 and 2017. Germany introduced the Network Enforcement Act to combat fake news and hate speech on social media in June 2017~\cite{loc}.

\section{Use Cases}
\label{sec:usecases}

The primary motivation to create this dataset was to study government censorship on social media. Here we provide possible use cases related to this topic, such as analysis of censorship policies and the effect of censorship. We also provide use cases with a non-censorship focus such as studying disinformation and hate speech. Note that this list is not exhaustive. 

\paragraph{Analysis of Censorship Policies} This dataset provides an excellent ground to study censorship policies of countries. Not only that, but it could also reveal where the platform policies and censorship policies align, given that some accounts are later removed by Twitter. 

\paragraph{Effect of Censorship} The dataset could be used to measure the effect of censorship on censored users' behavior. Do users forgo using their accounts after being censored, or does the censorship backfire? Furthermore, what is the effect of censorship on other users, e.g. does the public engage more with censored tweets? Studies tackling these questions will shed light on the effect of censorship.

\paragraph{Hate Speech Detection} Hate speech detection lacks ground truth of hate speech as hate speech is rare and often removed by Twitter before researchers collect them, e.g. ~\cite{davidson2017automated} sampled tweets based on a hate-oriented lexicon and found only 5\% to contain hate speech. This dataset would help to build a hate speech dataset as many of the tweets censored by Germany and France appear to involve hate speech. Researchers can collect those tweets, annotate whether they include hate speech, and build a new dataset of tweets including hate speech. Such a dataset might be useful for hate speech detection in French and German. It would also be used to evaluate existing hate speech detection methods.

\paragraph{News and Disinformation} Some users who were censored appear to be dissidents actively campaigning against the governments censoring them. Such users propagate news and claims at high rates. It would be interesting to see which news pieces and claims are censored by the government, but not removed by Twitter. It would also be interesting to see what portion of these news stories were fake news and which claims were debunked.

\section{Caveats}
\label{sec:ethics}

\subsection{Limitations}

\subsubsection{Sampling Bias}
We collected our dataset from an extensive dataset which consists of a 1\% random sample of all tweets. Due to the nature of random sampling, we assume the latter dataset to be unbiased. However, some tweets in the random sample are retweets of other tweets. The inclusion of tweets that are retweeted makes the dataset biased towards popular tweets and users who are more likely to be retweeted. Biased datasets can impact studies that report who or what content is more likely to be censored or which countries censor more often. To overcome this issue, we also include an unbiased subset of the original dataset. This dataset is collected by mining the Twitter Stream Grab without collecting the retweets and the tweets they are retweeting. The dataset consists of 39,913 tweets. It is stored in \texttt{tweets\_debiased.csv}. 

\subsubsection{Depth}
Our dataset is mined from the 1\% sample of all tweets on Twitter. As we do not have access to the full sample, we acknowledge that this dataset is not exhaustive and advise researchers to take this fact into the account.

\subsubsection{Data Turnout}
Our dataset was mined from a live stream, i.e. at the time that the data was produced. However, Twitter or the users sometimes remove their data. Indeed, we found that 43.7\% of the tweets were removed. We acknowledge that the turnout rate for the censored tweets is low, but the amount of data that remains (328,783 tweets) is still significant. To address this, we provide the code to reproduce the process to mine the archive so that researchers can gain access to all tweets.

\subsection{Ethical Considerations}

Our dataset consists only of the tweet ids and the user ids. This ensures that the users who chose to delete their accounts or users who protected their accounts will not have their data exposed. Sharing only the ids also complies with Twitter's Terms of Service. Although our supplementary dataset is massive (16 million tweets), Twitter permits academic researchers to share an unlimited number of tweets and/or user ids for the sole purpose of non-commercial research~\cite{developer_agreement}.

We also acknowledge that, as with any dataset, there is the possibility of misuse. By observing our dataset, malicious third parties can learn which content is censored and, e.g., learn to counter the censor. However, we also note that our dataset is retrospective and not real-time. There is a span of at least six months between the publication of this dataset and the censorship of the user. Additionally, Twitter sends notifications to the accounts that are censored~\cite{twitter_withheld}, so our dataset does not unduly inform a user of their own censure. Censored users may deactivate their accounts to avoid their public data being collected. Another misuse could be that a government could use this dataset for a political objective, e.g. to automatically detect users they should censor. While this is indeed a concern, we believe that this is an issue that must be addressed through governance and not further data withholding. Additionally, the governments would have more resources (i.e. the list of users they censor) had they pursued such an objective.

\subsection{Compliance with FAIR Principles}
FAIR principles state that FAIR data has to be findable, accessible, interoperable, and reusable. Our dataset is findable, as we made it publicly available via Zenodo. By choosing Zenodo, we also ensure it is accessible by everyone who wishes to download it, regardless of university or industry status. The data is interoperable as it is in .csv format which can be operable by any system. Finally, it is reusable as long as Twitter and/or Internet Archive's Twitter Stream Grab exists. To further enhance its reusability, we share the necessary code for the reproduction of the dataset in a public Github repository.

\fontsize{9.0pt}{10.0pt} \selectfont
\bibliography{bib}
\bibliographystyle{aaai}

\end{document}